\documentclass[epsfig,onecolumn,floats,twocolumn,showpacs]{revtex4}
\usepackage{graphicx}
\usepackage{dcolumn}
\usepackage{bm}
\usepackage{amssymb}
\usepackage{amsmath}
\usepackage{epsfig}

\begin{document}

\title{Quantum phase transition of Bose-Einstein condensates on a ring nonlinear
lattice}
\author{Zheng-Wei Zhou$^{\text{1}}$}
\email{zwzhou@ustc.edu.cn}
\author{Shao-Liang Zhang$^{\text{1}}$}
\author{Xiang-Fa Zhou$^{\text{1}}$}
\author{Guang-Can Guo$^{\text{1}}$}
\author{Xingxiang Zhou$^{\text{1}}$}
\email{xizhou@ustc.edu.cn}
\author{Han Pu$^{\text{2}}$}
\email{hpu@rice.edu}
\address{$^{\text{1}}$Key Laboratory of Quantum
Information, University of Science and \\ Technology of China,
Hefei, Anhui 230026, P. R. China\\ $^{\text{2}}$Department of
Physics and Astronomy, and Rice Quantum Institute, Rice
University, Houston, Texas 77251-1892, USA}

\begin{abstract}
We study the phase transitions in a one dimensional Bose-Einstein
condensate on a ring whose atomic scattering length is modulated
periodically along the ring. By using a modified Bogoliubov method
to treat such a nonlinear lattice in the mean field approximation,
we find that the phase transitions are of different orders when the
modulation period is 2 and greater than 2. We further perform a full
quantum mechanical treatment based on the time-evolving block decimation
algorithm which confirms the mean field results and reveals interesting
quantum behavior of the system. Our studies yield important
knowledge of competing mechanisms behind the phase
transitions and the quantum nature of this system .

\end{abstract}


\maketitle

\section{Introduction}
In the past few years, ultracold atoms confined in optical lattices have
generated a great amount of excitement in the physics community. They provide
the unique opportunity to realize various many-body models that
are of fundamental importance in physics \cite{review}. More
recently, nonlinear lattices formed by periodically modulating atomic
interaction strengths have also received much attention. A
comprehensive review of nonlinear wave phenomena supported by nonlinear lattices
can be found in Ref.~\cite{malomed}. There are two basic physical systems
that can potentially realize nonlinear lattices, both of which can be
described by nonlinear Sch\"{o}dinger equations.
 One uses electromagnetic waves
subject to
inhomogeneous nonlinear optical media. The other one is based on atomic
Bose-Einstein condensates
(BECs) with modulated $s$-wave scattering length. In this work, we will focus on
the latter system although much of the physics are common to both.

We extend our previous work reported in Ref.~\cite{Qian} to study
a BEC on a one-dimensional (1D) nonlinear ring-shaped lattice. In
Ref.~\cite{Qian}, we considered a BEC on this ring lattice whose
atomic scattering length is modulated according to
\begin{equation}
 a(\theta) = a_0 \sin (d \theta),
\end{equation}
 where $\theta$ is the azimuthal angle along the ring and $d=2$ is the spatial
modulation frequency. We have shown that, as the modulation depth $a_0$ is
increased, the condensate can undergo
a second-order symmtry-breaking quantum phase transition from a soliton-like
state to a spatially periodic condensate that matches the
scattering length modulation. In the present work,
we generalize our investigation to a larger spatial
modulation frequency $d\ge 3$ and compare the results to the $d=2$ case.
We developed a new
mean-field technique to study the semi-classical behavior of the system, and
found that a similar symmetry-breaking phase transition occurs for $d\ge 3$ as the
modulation depth is increased. However, the phase transition is now of {\em
first order}. We also carried out a numerical full quantum mechanical treatment of the
system based on the time-evolving block decimation (TEBD) algorithm. Both static and dynamical properties of the
system are investigated.

\section{The Model Hamiltonian}
The system considered here is similar to that studied in
Ref.~\cite{Qian,Kanamoto}. $N$ bosons are confined in a toroid of
radius $R$ and cross sectional area $S$. By sufficiently
tightening the radial confinement and freezing the atoms in that
direction, we can treat the atoms as a one dimension system on a
ring. The Hamiltonian of the system can be written in the
following dimensionless form:
\begin{equation}
H=\int_0^{2\pi }d\theta \left[ -\widehat{\psi }^{\dagger }\left( \theta
\right) \frac{\partial ^2}{\partial \theta ^2}\widehat{\psi }\left( \theta
\right) +\frac{U }2\widehat{\psi }^{\dagger }\left(
\theta \right) \widehat{\psi }^{\dagger }\left( \theta \right) \widehat{\psi
}\left( \theta \right) \widehat{\psi }\left( \theta \right) \right] ,
\end{equation}
where the first term in the integral represents the kinetic energy and the
second the interaction energy. For simplicity in notations, we measure energy
in units of $%
\hbar ^2/(2mR^2)$. The dimensionless interaction energy $U\left(
\theta \right) =8\pi a\left( \theta \right) R/S$, where $a\left(
\theta \right) $ is the periodically modulated $s$-wave scattering length. In
our work, we
consider the situation where the scattering length between atoms
is modulated along the ring with $d$ periods: $a\left( \theta
\right) =a_0\sin \left( d\theta \right) $. As in Ref.~\cite{Kanamoto}, we define
the dimensionless interaction strength as:
\begin{equation}
\gamma \left( \theta \right) \equiv -\frac{U\left( \theta \right)
N}{2\pi }=\gamma _0\sin (d\theta ),
\end{equation}
where $\gamma _0=-4a_0RN/S$ represents the modulation depth of
the interaction parameter $\gamma \left( \theta \right)$.

By taking the Fourier expansion of the field operator
$\widehat{\psi }\left( \theta \right) =\sum_k\frac 1{\sqrt{2\pi
}}e^{ik\theta }\widehat{a}_k$, where $k$ takes integer values in
order to satisfy the periodic boundary condition and ${
\widehat{a}_k }$ is the bosonic annihilation operator for plane
wave mode with wavenumber $k$, the Hamiltonian can be rewritten
as:
\begin{eqnarray*}
H &=& \sum_kk^2\widehat{a}_k^{\dagger }\widehat{a}_k+\frac{i\gamma _0
}{4N}\sum_{klmn}\widehat{a}_k^{\dagger }\widehat{a}_l^{\dagger
}\widehat{a}_m\widehat{a}_n ( \delta _{d+m+n-k-l} \\ && -\delta
_{-d+m+n-k-l}).
\end{eqnarray*}
Since the number of atoms is fixed, we have $N=\sum_l
\widehat{a}_l^{\dagger }\widehat{a}_l$. Because of this, the kinetic energy term in the
Hamiltonian
can also be written as $\frac{1}{N}\sum_{k,l}k^2 \widehat{a}_k^{\dagger
}\widehat{a}_k\widehat{a}_l^{\dagger
}\widehat{a}_l$.

\section{Mean-Field Treatment}
\label{MFL}
In this section, we first consider the mean-field solution valid for $N \gg 1$.
In this case, the kinetic energy term can be approximated as
\[ \frac{1}{N}\sum_{k,l}k^2 \widehat{a}_k^{\dagger
}\widehat{a}_k\widehat{a}_l^{\dagger }\widehat{a}_l \approx
\frac{1}{N}\sum_{k,l}k^2 \widehat{a}_k^{\dagger }
\widehat{a}_l^{\dagger } \widehat{a}_k \widehat{a}_l \,, \] and
the total Hamiltonian can thus be cast into a biquadratic form as:
\begin{equation}
\label{bi}
H=\frac{1}N\sum_{i,j,k,l}{\alpha _{ijkl}}\widehat{a}_i^{\dagger
}\widehat{a}_j^{\dagger }\widehat{a}_k\widehat{a}_l \,.
\end{equation}
We will now describe a modified Bogoliubov method we use to find the stationary
solution and the excitations of the system.

\subsection{Modified Bogoliubov Approach}
In the absence of atomic interactions, the ground state is quite
trivial: all the atoms occupy the zero-momentum mode
$\widehat{a}_0$. In the presence of interaction, this is no longer
true. However, we may conjecture that the system condenses into a
different ground mode $\widehat{\chi}_0 $. This mode
$\widehat{\chi}_0 $, together with other orthogonal modes
$\{\widehat{\chi}_i\}$'s that form a complete set, are related to
the $\{\widehat{a}_i\}$ modes through a unitary transformation
$U$:
\begin{equation}
\left( \chi _0,\chi _1,\chi _2,...\right) ^T=U \,\left(
a_0,a_1,a_{2},...\right) ^T \,. \label{uni}
\end{equation}
In terms of $\{\widehat{\chi}_i\}$ and $\{\widehat{\chi}_i^\dag \}$, the biquadratic
Hamiltonian in Eq.~(\ref{bi}) takes the following form:
\begin{widetext}
\begin{eqnarray}
H &= & \frac{c_0}N \,\widehat{\chi }_0^{\dagger }\widehat{\chi
}_0^{\dagger }\widehat{\chi }_0\widehat{\chi
}_0+ \left(\frac{\widehat{\chi }_0^{\dagger }\widehat{\chi }_0^{\dagger
}}N \sum_{k,l\neq 0}c_{kl}\widehat{\chi }_k\widehat{\chi
}_l+h.c. \right)+\frac{\widehat{\chi }_0^{\dagger }\widehat{\chi
}_0}N \,\sum_{k,l\neq 0}d_{kl}\widehat{\chi }_k^{\dagger
}\widehat{\chi }_l \nonumber \\
&&+\left(\frac{\widehat{\chi }_0^{\dagger }}N \,\sum_{k,l,m\neq
0}p_{klm}\widehat{\chi }_k^{\dagger }\widehat{\chi
}_l\widehat{\chi }_m +h.c. \right)+\frac{1}N \,\sum_{k,l,m,n\neq
0}q_{klmn}\widehat{\chi }_k^{\dagger }\widehat{\chi }_l^{\dagger
}\widehat{\chi }_m\widehat{\chi
}_n + \frac{1}N \,\sum_{k,l}r_{kl}\widehat{\chi }_k^{\dagger
}\widehat{\chi }_l \,. \label{Hchi}
\end{eqnarray}
\end{widetext}
This Hamiltonian can be simplified by a few considerations. First,
it is assumed that most atoms will be in the condensate mode
$\widehat{\chi}_0 $. Under the mean field approximation, operators
for the macroscopically occupied condensate mode are replaced by
$c$-numbers, i.e., $\widehat{\chi}_0, \widehat{\chi}_0^\dag
\rightarrow \sqrt{N} $. Since occupation numbers in other
$\widehat{\chi}_k $ modes are very small, we can drop terms involving 3
or more operators in $\widehat{\chi}_k$ and $\widehat{\chi}_k^\dag
$ for $k \neq 0$. After this exercise, we obtain the following effective
Hamiltonian up to second order in $\{\widehat{\chi}_k\}$ and
$\{\widehat{\chi}_k^\dag \}$:
\begin{equation}
H_{\rm eff} =c_0N+\left(\sum_{k,l\neq 0}c_{kl}\widehat{\chi }_k\widehat{\chi
}_l+h.c.\right)+\sum_{k,l\neq 0}d_{kl}\widehat{\chi }_k^{\dagger
}\widehat{\chi }_l \,. \label{heff}
\end{equation}

This effective Hamiltonian can be diagonalized by the Bogoliubov
transformation and the system's elementary excitaitons are
quasiparticles in nature. In order to investigate the stability of
the system, we need to analyze the energy spectrum of these
quasiparticle excitations. For this purpose, we should work with
the following grand canonical operator to acccount for the
conservation of atom numbers:
\begin{equation}
K=H_{\rm eff}-\mu N \,. \label{K}
\end{equation}
Here, $\mu$ is the chemical
potential and can be
calculated from the condensate energy $E=\left\langle H \right\rangle \approx
c_0 N$ as:
\begin{equation}
\mu =\frac{\partial E}{\partial N}=c_0+\frac{\partial
c_0}{\partial N}N \,. \label{che}
\end{equation}
Now, we may diagonalize the operator $K$ by
using the Bogoliubov transformation on $\{\widehat{\chi}_k\}$ and $\{\widehat{\chi}_k^\dag \}$
and obtain the excitation spectrum of
quasiparticles. (This will be elaborated on later.) If there is no
imaginary excitation frequencies, we claim that
the condensate
mode $\widehat{\chi}_0$ is dynamically stable.

In the above prescription,
the key step is to search for the appropriate unitary
transformation defined in Eq.~(\ref{uni}) that transforms Hamiltonian (\ref{bi})
into (\ref{Hchi}). Although there
are a great number of unknown parameters in the undetermined
unitary matrix $U$, further analysis shows that only the elements in the first
row of $U$ are necessary
for determining the form of the Hamiltonian (\ref{Hchi}).

To see this, we note that in order to transform Hamiltonian
(\ref{bi}) into (\ref{Hchi}) via the unitary matrix $U$, a
fundamental requirement is to maintain the biquadratic terms of
the operators $\left( \widehat{\chi }_0^{\dagger },\widehat{\chi
}_0 \right) $ and to eliminate all the cubic terms. To this end,
we may use the operators $\left( \widehat{\chi }_i^{\dagger
},\widehat{\chi }_j \right) $ to represent the operators $\left(
\widehat{a}_l^{\dagger },\widehat{a}_m \right) $ in Hamiltonian
(\ref{bi}) via:
\begin{equation*}
\widehat{a}_i=\sum_ju_{ji}^{*}\, \widehat{\chi} _j
\end{equation*}
where $u_{ij}$ is the matrix element of the unitary matrix $U$,
and we obtain the following two equations :
\begin{eqnarray}
c_0 &=& \sum_{i,j,k,l}\,{\alpha
_{ijkl}}\,u_{0i}u_{0j}u_{0k}^{*}u_{0l}^{*} \,, \label{1st} \\
0 &=&
\sum_{i,j,k,l}\,(\alpha _{ijkl}+\alpha _{ijlk})
\left[u_{0i}u_{0j}u_{0k}^{*}(a_l-u_{0l}^{*}\chi _0)
 \right] \,. \label{2nd}
\end{eqnarray}
Since $\chi_0=\sum_iu_{0i}a_i$, Eq.~(\ref{2nd}) can be
recast into a set of equations in operators $\left\{
a_i\right\} $ (the number of this set of equations depends on the cut-off
of the Bose modes) and can be solved by numerical method. Once the
representation of the operator $\chi_0$ is determined, the parameter
$c_0$ and the chemical potential $\mu$ can be obtained by
solving Eqs.~(\ref{1st}) and (\ref{che}), respectively.
Therefore, Eqs.~(\ref{1st}) and (\ref{2nd}) together represent an algebraic form
of the
Gross-Pitaevskii (GP) equation. In Appendix~\ref{a1}, we will show that they are indeed
equivalent to the ordinary GP equation. The main advantage of
Eqs.~(\ref{1st}) and (\ref{2nd}) is that, in principle, all the stationary
states (both dynamically stable and unstable ones) of the system can be found.
When more than one stable solutions are found, the one that is dynamically stable and with the lowest energy
will be identified as the ground state of the system. In contrast, with ordinary GP
equation, using imaginary time evolution method one can only find the dynamically stable
states, and often just the ground state. Therefore, Eqs.~(\ref{1st}) and (\ref{2nd})
are supeior for studying the phase transitions in our system, where information beyond the ground state is needed.

Once the condensate state is determined, the Bogoliubov spectrum of quasiparticle
excitations above it can be found by diagonalizing $K$
defined in Eq.~(\ref{K}) using the Bogoliubov transformation. The details are
described in Appendix~\ref{a2}.

\subsection{Mean-field quantum phase transition}

Using the modified
Bogoliubov method outlined in the previous section, in principle we can find
all the stationary states (not just the ground state) and their excitation
spectrum for any modulated
atomic interactions. This provides a more thorough picture of the
energy landscape of the system and deeper insights into possbile quantum phase transitions
induced when certain parameters are varied (in our case, the modulation depth of the
interaction strength $\gamma_0$).

Our goal is to study the mean field quantum phase transition for
different modulation period
$d$ as the modulation depth $\gamma_0$ is varied. We concentrate on the
low-energy states, meaning stationary states with energy
close to that of the ground state. We find that there are mainly two types of
stationary states in the low energy regime. One type
has a density profile matching the modulated period of the
scattering length. We refer to such states as symmetric states. The other type
features a density profile that spontaneously breaks the symmetry of the
modulation. We refer to such states as asymmetric states. The asymmetric states
is always $d$-fold degenerate
with the peak density located at $\frac{(1+4i)\pi }{2d}$ ($i=0,1,...d-1$), where
the local interaction energy $U(\theta)$ reaches the minimum.
Under the mean field treatment (MFT), the symmetry-breaking quantum phase
transition from
the symmetric type to the asymmetric type have been studied
for uniform scattering length \cite{Kanamoto} and for $d=2$
periodic scattering length \cite{Qian}. Here, by taking advantage
of the modified Bogoliubov method, we investigate the critical point
of the quantum phase transitions and the prime mechanism driving such
quantum phase transitions for arbitrary modulation period $d$. We
find that there is a fundamental difference between $d=2$ and $d>2$.

\begin{figure}[htb]
\epsfig{file=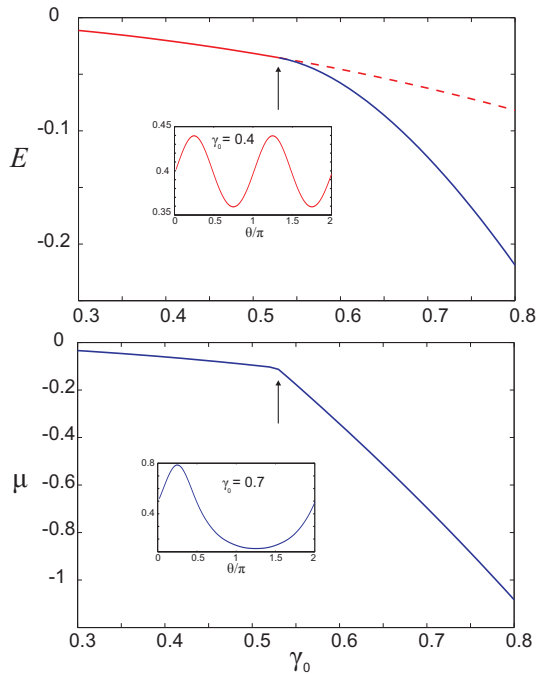,width=7cm}
\caption{(Color online) Upper panel: Energy
of the condensate for $d=2$. The
red solid line is for the dynamically stable symmetric state, the red dashed
line is for the dynamically unstable symmetric state and the blue solid line
is for the stable asymmetric state. Lower panel: Chemical potential of the
ground state. At the critical point $\gamma_0=0.528$ (indicated by arrows in the
plots) the ground state changes from the symmetric to the asymmetric type. Shown
in the insets are typical wave functions for symmetric and asymmetric states. }
\label{fig1}
\end{figure}

Figure \ref{fig1} shows the energies and chemcial potentials of low-energy Bose
condensate states by varying the parameter $\gamma _0$ for $d=2$. For small
$\gamma_0$, the ground state is a symmetric state. As $\gamma_0$ is increased, a
symmetry-breaking phase transition occurs at a critical value of
$\gamma_0=0.528$. At this point, the symmetric state becomes dynamically
unstable and the ground state changes to an asymmetric state. The ground state
chemcial potential shows a kink at this critical point, wheras the ground state
energy curve is smooth. This represents a second-order phase transition
\cite{Qian}. A similar behavior is also found in attractive BEC with unmodulated
scattering length \cite{Kanamoto}.

\begin{figure}[htb]
\epsfig{file=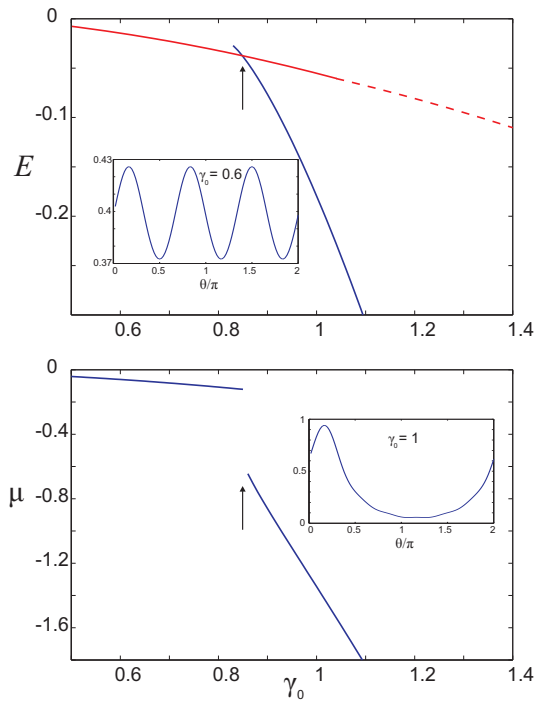,width=7cm}
\caption{(Color online) Same plots as in Fig.~\ref{fig1} for $d=3$. The
symmetry breaking phase transition occurs at the critical point
$\gamma_0=0.85$ (indicated by the arrows in the plots) and the dynamical
instability sets in for the symmetric state at $\gamma_0=1.04$. }
\label{fig2}
\end{figure}

Next, we turn our attention to the case of $d=3$. The energies and chemical
potentials as functions of $\gamma_0$ are plotted in Fig.~\ref{fig2}.
Similar to the $d=2$ case, for small $\gamma_0$, the ground state is
symmetric. As $\gamma_0$ is increased, a symmetry-breaking phase transition
occurs at a critical value of $\gamma_0=0.85$. However, unlike in the $d=2$
case, at this critical point, the symmetric state remains dynamically stable
although the energy of the asymmetric state drops below that of the symmetric
state. Furthermore, the ground state energy curve as a function of $\gamma_0$
shows a kink at this critical value, while the ground state chemical potential
becomes discontinuous at this point. Therefore, the phase transition at this point
is of first order. The dynamical instability of the symmetric
state does not occur until $\gamma_0=1.04$.

To summarize our mean-field results, we have found that the nature of
the phase transitions changes for different modulation period $d$ due to
the presence of competing mechanisms in this system.
When the modulation period $d=2$, the
symmetry-breaking phase transition is of second order and is induced by
dynamical instability of the associated states. For $d=3$, in contrast,
the symmetric-breaking phase transition is of first order and is driven by the
level crossing of different types of states. We have also investigated the cases
for $d=4$ and 5 and found similar behavior as in $d=3$.

\section{quantum mechanical treatment}

So far, we have limited our discussion to the mean-field approximation. Now we
perform a fully quantum mechanical examination of the system using a numerical method.
In our previous work \cite{Qian}, exact diagonalization is used for
this purpose. Here, we will use the TEBD algorithm \cite{Vidal}. This has a two-fold advantage compared to the exact diagonalization method: (1) It allows us to treat larger systems; and (2)
in addition to the
static properties,
we can also use the TEBD method to study the dynamical behavior of the system.

We first discretize the space by introducing an equidistant grid $\theta
_i=i\Delta \theta$,
$(i=0,1,...M-1)$. We then replace the field operator $\widehat{\psi }
\left( \theta_i \right)$ by $\widehat{\psi }_i /\sqrt{\Delta
\theta }$, where $\widehat{\psi}_i$ is a bosonic annihilation
operator. In doing so, integrals can be replaced by sums and the
second derivative in the kinetic energy term can be approximated by the
difference quotient $\frac{\partial ^2}{\partial \theta
^2}\widehat{\psi }\left( \theta _i\right) \approx \left[
\widehat{\psi }\left( \theta _{i+1}\right) +\widehat{\psi }\left(
\theta _{i-1}\right) -2\widehat{\psi }\left( \theta _i\right)
\right] /\Delta \theta ^2$. Finally, the discretized Hamiltonian
is  \cite{Michael1}:
\begin{widetext}
\begin{equation}
H=-\frac 1{\Delta \theta ^2}\sum_{i=0}^{M-1}\left( \widehat{\psi
}_i^{\dagger }\widehat{\psi }_{i+1}+h.c.\right) +\frac 1{\Delta
\theta ^2}\sum_{i=0}^{M-1}\widehat{\psi }_i^{\dagger
}\widehat{\psi }_i+\sum_{i=0}^{M-1}\frac{U_i}{2\Delta \theta
}\widehat{\psi }_i^{\dagger }\widehat{\psi }_i^{\dagger
}\widehat{\psi }_i\widehat{\psi }_i.
\end{equation}
\end{widetext}
Here, the periodic boundary condition leads to the relation
$\widehat{\psi }_0=\widehat{\psi }_M$. For TEBD algorithm with
periodic boundary condition, we refer to Ref.~\cite{Naidon}. In
our numerical treatment, we typically divide the ring into $M=60$
equidistant grids. Our code is adapted from the open source
package maintained by the group of Lincoln Carr \cite{open}.

\subsection{many-body ground-state energy}

\begin{figure}[htb]
\epsfig{file=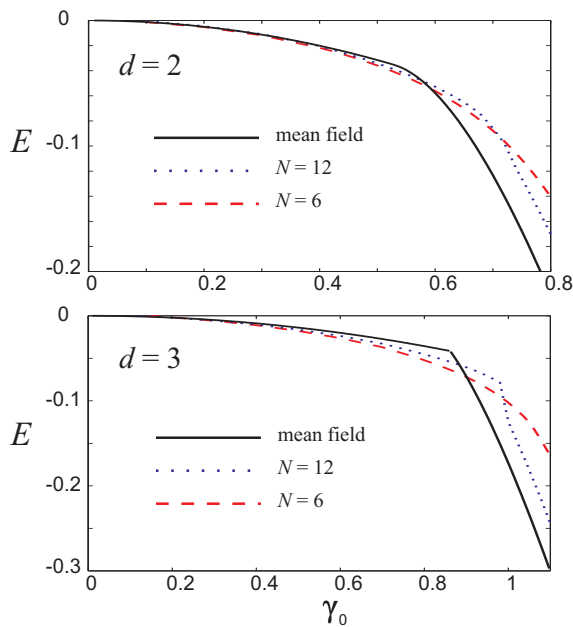,width=7.5cm,} \caption{(Color online) Many-body
ground-state energy per atom for $N=6,12$ and
modulation period $d=2$ (upper panel) and $d=3$ (lower panel). The black solid
lines are the mean-field results.}
\label{energy}
\end{figure}

In Fig.~\ref{energy}, we plot the ground-state energy per atom of
the many-body system as functions of $\gamma_0$ for the modulation
period $d=2$ and $d=3$ . We see that the ground-state energy curve
approaches the MFT result as $N$ increases, which is consistent
with the usual quantum-semiclassical crossover behavior for
finite-size quantum systems.

\subsection{quantum correlation}

In the quantum mechanical treatment, unlike the mean field results,
the spontaneous symmetry breaking of density distribution of
ground state wavefunction in real space dose not occur. The
density profile of the quantum mechanical ground state always matches the
spatial modulation of the scattering length \cite{Qian}. However, we
can still gain important insights into the change in the characteristics of the
wavefunctions by examining
the quantum correlation which is neglected in the mean-field study.

For further discussion, we define the partial number operator between
the interval $\theta
\in [\varphi_i, \varphi_f]$ as:
\begin{equation}
\widehat{n}_{\left[ \varphi _i,\varphi _f\right] }=\frac 1{2\pi
}\int_{\varphi _i}^{\varphi _f}d\theta \,\widehat{\psi }^{\dagger
}\left( \theta \right) \widehat{\psi }\left( \theta \right) .
\end{equation}

\begin{figure}[htb]
\epsfig{file=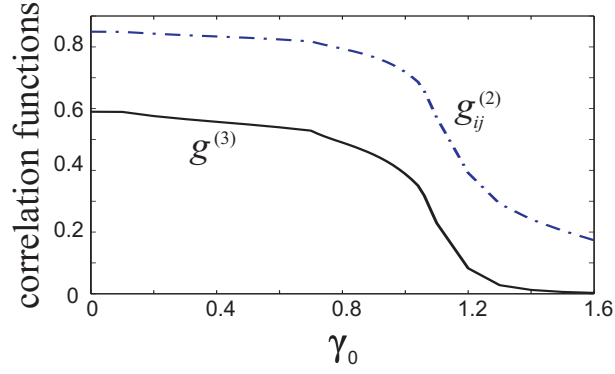,width=8cm,} \caption{(Color
online) Bipartite correlation
$g_{ij}^{\left( 2\right) }$ and tripartite correlation $g^{\left(
3\right) }$ as functions of $\gamma_0$ for $N=6$ and $d=3$.}
\label{corr}
\end{figure}

For the particular case of $d=3$, we
define three partial particle number operators as follows:
\begin{equation}
\widehat{n}_i=\widehat{n}_{\left[ (i-1)\frac{2\pi }3,i\frac{2\pi
}3\right] }, \quad i=1,2,3.
\end{equation}
Using these number operators, we can define the bipartite and tripartite correlation
functions as:
\begin{eqnarray*}
g_{ij}^{\left( 2\right) } &=& \frac{\left\langle
\widehat{n}_i\widehat{n}_j\right\rangle }{\left\langle
\widehat{n}_i\right\rangle \left\langle \widehat{n}_j\right\rangle
} \,,
\\
g^{\left( 3\right) } &=& \frac{\left\langle
\widehat{n}_1\widehat{n}_2\widehat{n}_3\right\rangle
}{\left\langle \widehat{n}_1\right\rangle \left\langle
\widehat{n}_2\right\rangle \left\langle \widehat{n}_3\right\rangle
}\,.
\end{eqnarray*}

We plot the bipartite and tripartite correlations as functions of $\gamma_0$
in Fig.~\ref{corr}. All these correlations are monotonically decreasing
functions of $\gamma_0$. In our calculation, the three two-body correlation
functions $g_{12}^{(2)}$,  $g_{13}^{(2)}$ and  $g_{23}^{(2)}$ are essentially
identical, which is also expected from the symmetry of the system. At
$\gamma_0=0$,
the ground state is exactly known: $\left|
\Psi \right\rangle _{ground}=\frac{\left( a_0^{\dagger }\right)
^N}{\sqrt{N!}}\left|{\rm vac}\right\rangle $. The theoretical values of
bipartite and tripartite correlations can be obtained as:
\begin{eqnarray*}
g_{ij}^{\left( 2\right) }\left( \gamma _0=0\right) &=&1-\frac 6{\pi
^2N}  \sum_{l=1}^\infty \frac 1{l^2}  \,,\\
g^{\left( 3\right) }\left( \gamma _0=0\right) & =& 1-\frac{18}{\pi
^2N} \sum_{l=1}^\infty \frac 1{l^2}  +O\left( \frac
1{N^2}\right)\,,
\end{eqnarray*}
which are in good agreement with the numerical
results. When $N$ goes to infinity, all the bipartite and
tripartite correlations approach unity at $\gamma_0=0$.

Figure~\ref{corr} shows that although both bipartite and
tripartite correlations decay as the interaction parameter
$\gamma_0$ increases, the tirpartitle correlation $g^{\left(
3\right) }$ decays into zero much faster than the bipartitle
correlation $g_{ij}^{\left( 2\right) }$. For the case of $N=6$ as
illustrated in Fig.~\ref{corr}, $g^{\left( 3\right) }$ is
essentially zero at $\gamma_0=1.6$, while all the $g_{ij}^{\left(
2\right) }$ are clearly non-zero at the same $\gamma_0$. This is
reminiscent of the three-body entangled $W$-state, which can be
written as \[ |W \rangle = \frac{1}{\sqrt{3}}\, \left(|100 \rangle
+ |010 \rangle + |001 \rangle \right) \,.\] For the $W$-state, the
tripartite entanglement characterized by the 3-tangle disappears
and the bipartite entanglement characterized by concurrence
remains finite \cite{Coffman}. For large $\gamma_0$, our
mean-field treatment presented earlier reveals that the ground
state is characterized by asysmmetric state with three-fold
degeneracy. Each of these degenerate mean-field state features a
density peak at $\theta = (1+4i\pi)/6$ ($i=1,2,3$). In the quantum
treatment, this degeneracy is lifted by quantum fluctuations, and
the non-degenerate quantum ground state may be regarded as roughly
a $W$-state formed by these three mean-field states.

\subsection{Single-particle density matrix}
Another important quantity to characterize the many-body state is the
single-particle density matrix  $\rho ^{\left( 1\right)
}$ whose matrix element is defined as \cite{Penrose,Leggett}:
\begin{equation}
\rho _{ij}^{\left( 1\right) }=\left\langle \widehat{\psi
}_i^{\dagger }\widehat{\psi }_j\right\rangle \,,
\end{equation}
where the expectation value is calculated with respect to the ground state
obtained using the TEBD method. Roughly speaking, $\rho _{ij}^{\left( 1\right)
}$ represents the probability amplitude of finding one particle at site $i$ and
at the same time another particle at site $j$.

\begin{figure}[htb]
\epsfig{file=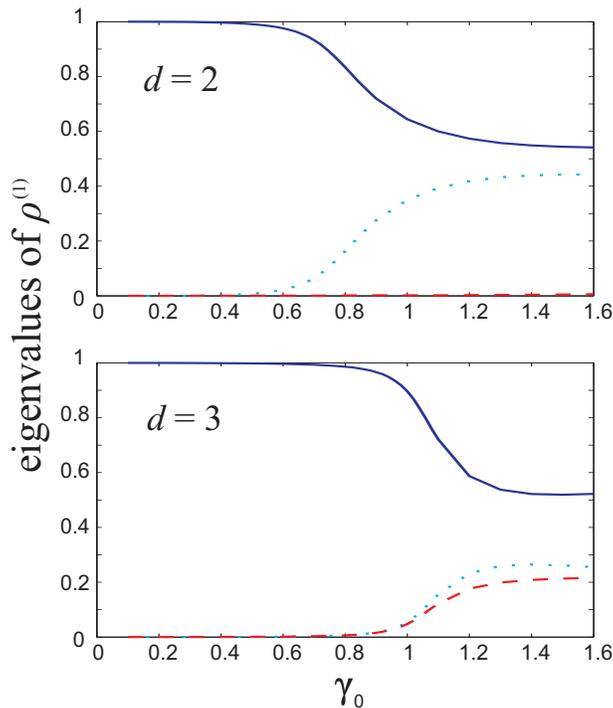,width=8cm,} \caption{(Color
online) The largest three eigenvalues of the single-particle density matrix
$\rho^{(1)}$ for $N=6$.}
\label{fig5}
\end{figure}

Because the matrix $\rho ^{\left( 1\right) }$ is Hermitian it can
be diagonalized as:
\begin{equation}
\rho ^{\left( 1\right) }=N\sum_{i} p_i \,\varphi _i^{*}\varphi _i \,,
\end{equation}
where the eigenvalues $p_i$ are non-negative and satisfy the
constraint $\sum_i p_i=1$. For a bosonic system with $N \gg 1$,
$p_i$ is closely related to the condensate fraction of the system. It is easy to see that, the system will be in a simple
condensate state if and only if the largest eigenvaule is of order
unity and all the other $p_{i}$'s are of order $O(N^{-1})$. If
there are multiple $p_i$'s of order unity, the system is said to
be in a fragmented condensate because all the corresponding states
have appreciable occupation numbers. Finally, if all the $p_i$'s
are of order $O(N^{-1})$, then the system is not Bose condensed.
It is reasonable to speculate that, at small $N$, the condensate
fraction versus the strength parameter $\gamma_0$ exhibits quantum
crossover behavior. For $N=6$ with the modulation periods $d=2$
and $d=3$, we plot the three largest $p_i$'s versus $\gamma_0$ in
Fig.~\ref{fig5}. The figure shows that at small $\gamma_0$, the
system may be characterized as a simple condensate. It becomes
more and more fragmented as $\gamma_0$ is increased. In the large
$\gamma_0$ limit, the eigenvalues approach some steady state
values and there are $d$ eigenvalues which are much larger than
the rest. This is consistent with the earlier argument that at
large $\gamma_0$, the quantum many-body ground state can be
roughly regarded as superpositions of the $d$-fold degenerate
mean-field ground states.

\begin{figure}[htb]
\epsfig{file=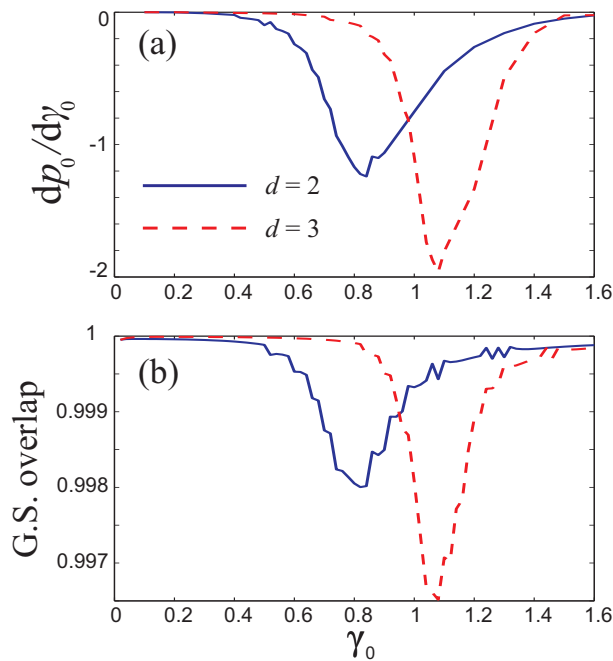,width=8cm,} \caption{(Color
online) (a) Derivative of the largest single-particle density matrix
eigenvalue with respect to the interaction strength modulation depth and (b)
overlap of the ground-state wave function versus $\gamma_0$ for
the particle number $N=6$.}
\label{fig6}
\end{figure}

To further quantify the crossover from a simple condensate to a fragmented
condensate, we
adopt two methods as described below. In Fig.~\ref{fig6}(a), we plot
$dp_0/d\gamma_0$ as a function of $\gamma_0$, where $p_0$ is the largest
single-particle density matrix eigenvalue. In Fig.~\ref{fig6}(b), we plot the
overlap of the ground-state wave function \cite{Zanardi} $|\langle G(\gamma_0)|
G(
\gamma _0+\delta \gamma) \rangle |$ for a small value of $\delta \gamma=0.02$,
where $|G(\gamma_0) \rangle$ denotes the ground state at $\gamma_0$. Both of
these quantities measure how fast the characteristics of the ground state change as
$\gamma_0$ is varied. These two measures provide consistent results: both
exhibit a dip at some critical value
 $\gamma_0=0.84$ for $d=2$ and $\gamma_0=1.08$
for $d=3$, which we may define as the critical modulation depth where the system
crosses from a simple condensate to a fragmented condensate.

\subsection{Time evolution of the survival probability}

Up to now, we have focused our attention on the static properties of the ground
state. The low-energy excitations
of the system in the vicinity of the crossover are also important
because they expose fine characteristics valuable for understanding the system
dynamics
in the crossover. A powerful tool to study this is the time evolution of
ground state survival probability  \cite{Felker,Wang}. It describes the
dynamical behavior
of the system's ground state under small perturbation in parameters in the
system Hamiltonian.
In our problem, we parameterize the Hamiltonian using the interaction strength
modulation depth $\gamma_0$ such that $H=H(\gamma_0)$. If $\gamma_0$ has a small
variation $\delta \gamma$ so
that $\gamma_0 \rightarrow \gamma_0+\delta \gamma$, the ground state's survival
probablity
is defined as
\begin{equation}
M(t)=\left| \left\langle G\left( \gamma _0\right) \right|
\exp\{-iH\left( \gamma _0+\delta \gamma \right) t\}\left| G\left(
\gamma _0 \right)\right\rangle \right| ^2 \,.
\label{M}
\end{equation}
The survival probability can be considered as a special case
of the quantum Loschmidt echo \cite{Peres}. Incidentally, the numerical method
we used to simulate our system, the TEBD algorithm, is very convenient in
calculating
the system's time evolution and hence the survival probability.

\begin{figure}[htb]
\epsfig{file=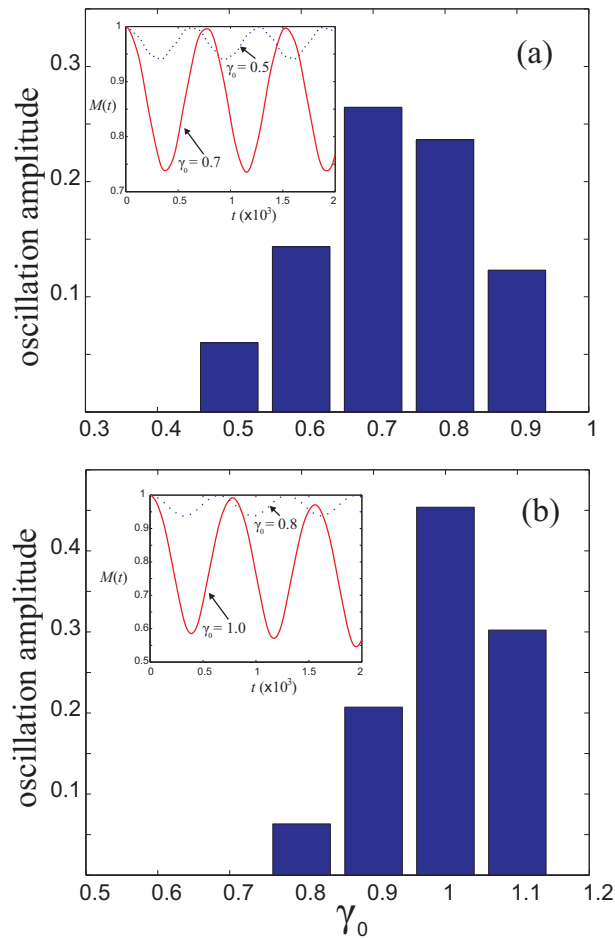,width=8cm,} \caption{(Color online) The amplitude of the
oscillation of the survival probability $M(t)$ for (a) $d=2$ and (b) $d=3$. The
inset shows the dyanmics of $M(t)$ for different values of $\gamma_0$.}
\label{fig7}
\end{figure}

We calculated the ground state survival probability $M(t)$ and
plot the results in Fig.~\ref{fig7} for $N=6$ and the perturbation in
$\gamma_0$ is
$\delta\gamma = 0.1$. $M(t)$ exhibits roughly sinusoidal oscillations in time.
The amplitude of the modulation reaches the maximum near critical points which
are consistent with those found in previous static study and shown
in Fig.~\ref{fig6}. Indeed, the curve for the
oscillation amplitude of
the ground state's survival probability can be used to predict the overlap
between the perturbed
and original ground state wave functions which is plotted in Fig.~\ref{fig6}(b).
The larger the oscillation amplitude in $M(t)$, the more
sensitive the ground state wavefunction to the perturbation in the Hamiltonian.

\section{Summary}
In conclusion, we have made a systematic investigation of a
condensate on a nonlinear ring lattice. Our studies show that the
properties of the system are sensitive to the modulation period
$d$ of the interaction strength. In particular, the meanfield
symmetry breaking phase transition is second order when the
modulation period $d$ is 2 but first order when $d > 2$, due to
competing mechanisms present in the system driving these
transitions. Our full quantum mechanical treatment based on the
TEBD method reveals the behavior of many important quantities that
are essential to the characterization of the system physics.

That the mean-field symmetry breaking phase transition changes
from second to first order when $d$ changes from 2 to larger than
2 is somewhat surprising. The change of the order of the phase
transition may be related to the change of the length scale
associated with the modulation of the scattering length: For $d >
2$, this length scale is smaller compared with $d = 2$. Recently,
Mayteevarunyoo et al. studied the symmetry breaking transition in
a BEC subject to a nonlinear double-well potential and found that
the width of the nonlinear potential plays an important role in
controlling the transition \cite{Mayteevarunyoo}, a phenomenon
that may be related to what we discovered in the current work.
This is certainly one of the peculiar properties of nonlinear
potentials that deserves further investigation.

\section{Acknowledgments} This work was funded by NFRP 2011CB921204
, and NNSF (Grant Nos. 60921091, 10874170, 10875110). Z. -W. Zhou
gratefully acknowledges the support of the K. C. Wong Education
Foundation, Hong Kong. HP acknowledges support from U.S. NSF. The
authors thank Yong-Jian Han, Biao Wu, Shi-Liang Zhu, Hui Zhai,
Wu-Ming Liu and Wen-Ge Wang for helpful discussions and comments.
Z. -W. Zhou appreciates the hospitality of the Kavli Institute of
Theoretical Physics in Beijing, where part of this work was
completed.

\appendix
\section{Proof of the equivalence between the algebraic type
and time-independent Gross-Pitaevskii equation}
\label{a1}

Following the argument in Sec.~I, we obtain the Hamiltonian in the limit $N \gg
1$ as
\begin{eqnarray}
H &=& \sum_{kl}\frac{k^2}{N} a_k^{\dagger }a_l^{\dagger }a_k
a_l+\frac{i\gamma _0}{4N}\sum_{klmn}a_k^{\dagger }a_l^{\dagger
}a_ma_n\left( \delta _{d+m+n-k-l} \right.  \nonumber \\
&& \;\left.  -\delta _{-d+m+n-k-l}\right)  \,,
\label{A2}
\end{eqnarray}
which can be written in a simplified form:
\begin{equation}
H=\frac{1}{N}\sum_{ijkl} {\alpha _{ijkl}}a_i^{\dagger }a_j^{\dagger
}a_ka_l \,.
\end{equation}
In Sec.~\ref{MFL}, we obtain the
algebraic type GP equations (\ref{1st}) and (\ref{2nd}) which we rewrite here
as:
\begin{eqnarray}
c_0 &=& \sum_{ijkl}\,{\alpha
_{ijkl}}\,u_{0i}u_{0j}u_{0k}^{*}u_{0l}^{*} \,, \label{A4} \\
0 &=&
\sum_{ijkl}\,(\alpha _{ijkl}+\alpha _{ijlk})
\left[u_{0i}u_{0j}u_{0k}^{*}(a_l-u_{0l}^{*}\chi _0)
 \right] \,. \label{A5}
\end{eqnarray}
Here, there is the unitary relation: $\sum_{l}u_{0l}^{*}u_{0l}=1$.
Based on Eqs.~(\ref{A2}) and (\ref{A4}), we have:
\begin{equation}
c_0=\sum_{i,j,k,l}\alpha
_{ijkl}u_{0i}u_{0j}u_{0k}^{*}u_{0l}^{*}=A+B,
\end{equation}
where
$A=\sum_{kl}k^2u_{0k}u_{0l}u_{0k}^{*}u_{0l}^{*}=\sum_{k}k^2u_{0k}u_{0k}^{*}$
and $B=\frac{i\gamma
_0}4\sum_{klmn}u_{0k}u_{0l}u_{0m}^{*}u_{0n}^{*}\left( \delta
_{d+m+n-k-l}-\delta _{-d+m+n-k-l}\right) $. Furthermore, by
considering Eq.~(\ref{A5}), the following relation can be
derived:
\begin{widetext}
\begin{equation}
\sum_{kl}k^2u_{0k}u_{0l}(u_{0k}^{*}a_l+u_{0l}^{*}a_k)+\frac{i\gamma
_0}4\sum_{klmn}u_{0k}u_{0l}(u_{0m}^{*}a_n+u_{0n}^{*}a_m)\left(
\delta _{d+m+n-k-l}-\delta _{-d+m+n-k-l}\right) =2c_0\chi _0 \,.
\label{A7}
\end{equation}
By taking advantage of the unitary relation:
$\chi_0=\sum_lu_{0l}a_l$, Eq.~(\ref{A7}) can be decomposed into
the following algebraic equations depending on the boson operator
$a_l$:
\begin{equation}
l^2u_{0l}+\frac{i\gamma _0}2\sum_{kmn}u_{0k}u_{0n}u_{0m}^{*}\left(
\delta _{d+m+l-k-n}-\delta _{-d+m+l-k-n}\right) =\mu u_{0l} \,,
\label{A8}
\end{equation}
where $\mu =2c_0-A=c_0+\frac{\partial c_0}{\partial N}N$.
\end{widetext}

The time-independent GP equation describing Bose gases on a ring
with periodic scattering length is:
\begin{equation}
-\frac{\partial ^2}{\partial \theta ^2}\psi \left( \theta \right)
-2\pi \gamma _0\sin (d\theta )\left| \psi \left( \theta \right)
\right| ^2\psi \left( \theta \right) =\mu \psi \left( \theta
\right) \,,
\label{A9}
\end{equation}
with the boundary condition:
$
\psi \left( \theta \right)=\psi \left( 2\pi+\theta \right)
$.
By replacing $\psi \left( \theta \right) =\frac 1{\sqrt{2\pi
}}\sum_lu_{0l}e^{-il\theta }$ into Eq.~(\ref{A9}), the
algebraic equations same as Eq.~(\ref{A8}) can be derived. This demonstrates the
equivalence between Eqs.~(\ref{1st}) and (\ref{2nd}) and the usual GP equation.

\section{Finding Excitation Spectrum}
\label{a2}
The Bogoliubov spectrum of quasiparticle excitations can be
determined by diagonalizing operator $K$ as defined in Eq.~(\ref{K}). However,
here, we can not obtain
the representation of Eq.~(\ref{K}) using the modes $\left\{
\chi_l^\dagger, \chi_m\right\} $ because the unitary
transformation $U$ is unknown except for its matrix elements in the
first row. This difficulty can be overcome by representing the last two terms at
the r.h.s. of
Eq.~(\ref{K}) using the original mode operators $\left\{ a_l^\dagger,
a_m\right\} $:
\begin{equation}
K=(c_0-\mu )N+\sum_{mn}A_{mn}a_m^{\dagger
}a_n+\sum_{mn}(B_{mn}a_ma_n+h.c.) \,, \label{b1}
\end{equation}
where $\sum_{mn}B_{mn}a_ma_n=\sum_{i,j,k,l}\alpha _{ijkl} \left\langle
a_i^{\dagger }a_j^{\dagger }\right\rangle :a_ka_l:$ and
\begin{widetext}
\[\sum_{mn}A_{mn}a_m^{\dagger }a_n=\sum_{i,j,k,l}\alpha
_{ijkl}\left\{ \left\langle a_i^{\dagger }a_k\right\rangle
:a_j^{\dagger }a_l: +\left\langle a_j^{\dagger }a_l\right\rangle
:a_i^{\dagger }a_k:+\left\langle a_i^{\dagger }a_l\right\rangle
:a_j^{\dagger }a_k:+\left\langle a_j^{\dagger }a_k\right\rangle
:a_i^{\dagger }a_l:\right\}-\mu (\sum_na_n^{\dagger }a_n-N)\,.\]
\end{widetext}
Here  $\left\langle {A}\right\rangle $ refers to the amplitude of
the operator $A$ projecting onto the operators
$\chi_0^\dagger\chi_0$, $\chi_0^\dagger\chi_0^\dagger$, or
$\chi_0\chi_0$ and $:A:$ refers to the corresponding component in
the operator $A$ orthogonal to the operators
($\chi_0^\dagger\chi_0$, $\chi_0^\dagger\chi_0^\dagger$,
$\chi_0\chi_0$). For instance,
\begin{equation*}
\left\langle a_i^{\dagger }a_j^{\dagger }\right\rangle
:a_ka_l:=u_{0i}u_{0j}\left( a_k-u_{0k}^{*}\chi _0\right) \left(
a_l-u_{0l}^{*}\chi _0\right).
\end{equation*}

To find
the energy spectrum for quasiparticle excitation, a generalized
Bogoliubov method ( see reference \cite{Milstein}) can be used to
diagonalize Eq.~(\ref{b1}). Briefly, by
introducing the row and column vectors
\begin{equation}
\varsigma =\binom a{a^{\dagger }},\quad \varsigma ^{\dagger }=\left(
a^{\dagger }a\right)
\end{equation}
and defining the matrix
\begin{equation}
M=\left(
\begin{array}{ccc}
A & 2B \\
2B^{*} & A^{*}
\end{array}
\right)
\end{equation}
we may rewrite the operator $K$ as:
\begin{equation}
K=(c_0-\mu )N+ \frac 12\varsigma ^{\dagger }M\varsigma -\frac 12 {\rm Tr}A \,.
\end{equation}
Here, the matrix $A=\left[ A_{mn}\right] $ and the matrix
$B=\left[ B_{mn}\right] $. Now $K$ can be
diagonalized by introducing the appropriate canonical
transformation $T$: $\beta =T\varsigma $. Finally, the operator
$K$ has the diagonal form:
\begin{equation}
K=(c_0-\mu )N+\frac 12\beta ^{\dagger }\eta T\eta MT^{-1}\beta -\frac 12 {\rm
Tr}A \,,
\end{equation}
where the matrix $\eta =I\oplus (-I)$ and $T\eta MT^{-1}$ is a
diagonal matrix.

\end{document}